\documentclass[aps,prl,twocolumn,showpacs,floatfix]{revtex4}

\usepackage{dcolumn}
\usepackage{amssymb}
\usepackage{graphicx}

\begin{document}

\newcommand \be {\begin{equation}}
\newcommand \ee {\end{equation}}
\newcommand \bea {\begin{eqnarray}}
\newcommand \eea {\end{eqnarray}}
\newcommand \nn {\nonumber}
\newcommand \la {\langle}
\newcommand \rl {\rangle_L}

\title{Boltzmann and hydrodynamic description for self-propelled particles}
\author{Eric Bertin,$^1$ Michel Droz,$^1$ and Guillaume Gr\'egoire$^2$}
\affiliation{$^1$ Department of Theoretical Physics, University of Geneva,
CH-1211 Geneva 4, Switzerland\\
$^2$ Mati\`ere et Syst\`emes Complexes, UMR 7057, CNRS-Universit\'e
Paris 7, F-75251 Paris Cedex 05, France}

\date{\today}

\begin{abstract}
We study analytically the emergence of spontaneous collective motion
within large bidimensional groups of self-propelled particles with noisy
local interactions, a schematic model for assemblies of biological organisms.
As a central result, we derive from the individual dynamics
the hydrodynamic equations for the density and velocity fields,
thus giving a microscopic foundation to the phenomenological equations
used in previous approaches.
A homogeneous spontaneous motion emerges below a transition line in the
noise-density plane. Yet, this state is shown to be unstable against spatial
perturbations, suggesting that more complicated structures
should eventually appear.

\end{abstract}


\maketitle

Collective motion of self-propelled interacting agents has become
in recent years an important topic of interest for statistical physicists.
Phenomena ranging from animal flocks (e.g.~fish schools or bird
flocks) \cite{Parrish}, to bacteria colonies
\cite{Bonner}, human crowds \cite{Helbing}, molecular motors
\cite{Harada}, or even interacting robots \cite{Sugawara},
depend only on a few general properties of the interacting
agents \cite{physbio,AnnPhys}.
From a physicist viewpoint, it is thus of primary importance to
analyze generic minimal models that could capture the emergence of collective
motion, without entering the details of the dynamics of each particular
system.
In this spirit, Vicsek \textit{et~al.} proposed a simple model
\cite{Vicsek}, defined on a continuous plane,
where ``animals'' are represented schematically as point
particles with a velocity vector of constant magnitude.  Noisy
interaction rules tend to align the velocity of any given particle with
its neighbors. A continuous transition from a disordered state at high
enough noise to a state where a collective motion arises was found
numerically \cite{Vicsek}. Recent numerical simulations confirmed the
existence of the transition, and suggested that the transition may be
discontinuous, with strong finite size effects \cite{Gregoire03,Gregoire04}.
In other approaches, velocity vectors have been associated with classical
spins \cite{Csahok95,Huepe}; lattice Boltzmann models have also been
proposed \cite{Bussemaker}.

However, apart from this large amount of numerical data, little
analytical results are available.  Some coarse-grained descriptions
of the dynamics in terms of phenomenological hydrodynamic equations
have been proposed \cite{Toner,Ramaswamy,AnnPhys,Csahok97},
on the basis of symmetry and conservation laws arguments.
Accordingly, the coefficients entering these equations have no microscopic
content, and their dependence upon external parameters is unknown.
Renormalization group analysis \cite{Toner} and numerical studies
\cite{Csahok97} confirm the presence of a nonequilibrium phase transition
in such systems. 
Still, a first-principle analytical approach based on the dynamics of
individuals on a continuous space is, to our knowledge, still lacking.
Such an approach would be desirable to gain a better
understanding of the spontaneous symmetry
breaking in two-dimensional systems with continuous rotational
symmetry, a phenomenon that cannot occur in equilibrium systems
due to the presence of long wavelength modes, as shown
by Mermin and Wagner \cite{Wagner}. Indeed, although the Mermin-Wagner
theorem does not hold in nonequilibrium system, one may wonder
whether long wavelength modes still play an important role \cite{Toner}.

In this short note, we introduce a microscopic bidimensional model of
self-propelled particles with noisy and local interaction rules
tending to align the velocities of the particles.
We derive analytically hydrodynamic equations for the density and velocity
fields, within a Boltzmann approach.
The obtained equations are consistent with previous
phenomenological proposals \cite{Toner,Ramaswamy,AnnPhys,Csahok97}.
Most importantly, we obtain explicit expressions for the coefficients
of these equations as a function of the microscopic parameters.
This allows us to analyze the phase diagram of the model
in the noise-density plane.

\begin{figure}[b]
\centering\includegraphics[width=8.3cm,clip]{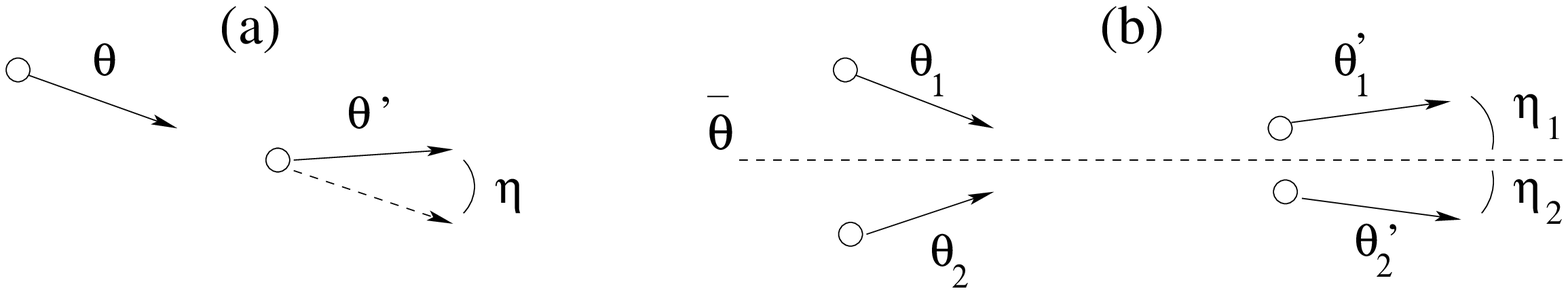}
\caption{\sl Schematic view of the dynamics of the model: (a) self-diffusion
events, (b) binary collisions with alignment interactions
--see text for notations.
}
\label{fig-model}
\end{figure}

\textit{Definition of the model.}--
We consider self-propelled point-like particles moving on a
continuous plane, with a velocity vector $\mathbf{v}$
of fixed magnitude $v_0$ (to be chosen as the velocity unit) in a reference
frame --hence, Galilean invariance no longer holds.
The velocity of the particles is simply defined by the angle $\theta$
between $\mathbf{v}$ and an arbitrary reference direction.
Particles evolve ballistically until they experience either a self-diffusion
event (a random ``kick''), or a binary collision that tends to align
the velocities of the two particles.
To be more specific, the velocity angle $\theta$ of any particle is
changed with a probability $\lambda$ per unit time to a value
$\theta'=\theta+\eta$ [Fig.~\ref{fig-model}(a)], where $\eta$ is a
Gaussian noise with distribution $p_0(\eta)$ and variance $\sigma_0^2$.
In addition, binary collisions occur when the distance between
two particles becomes less than $d_0$ (in the following, we set
$d_0=\frac{1}{2}$).
The velocity angles $\theta_1$ and $\theta_2$ of
the two particles are then changed into
$\theta_1'=\overline{\theta}+\eta_1$ and $\theta_2'=\overline{\theta}+\eta_2$
[Fig.~\ref{fig-model}(b)], where
$\overline{\theta}=\text{Arg}(e^{i\theta_1}+e^{i\theta_2})$
is the average angle,
and $\eta_1$ and $\eta_2$ are independent Gaussian noises with the same
distribution $p(\eta)$ and variance $\sigma^2$, that may differ from
$\sigma_0^2$.

\begin{figure}[t]
\centering\includegraphics[width=7.5cm,clip]{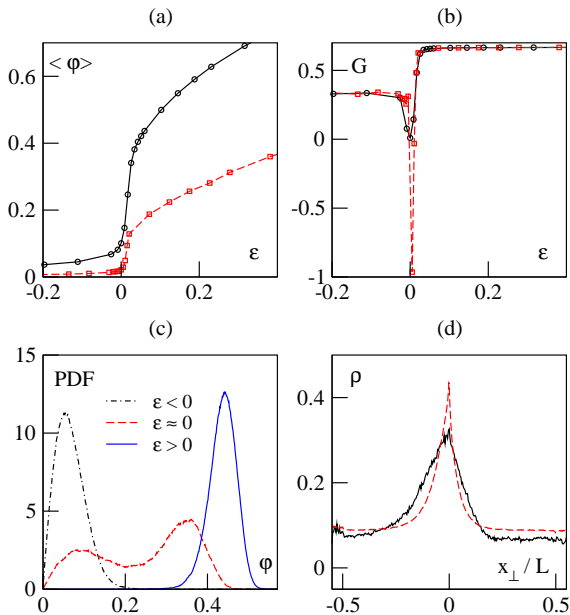}
\caption{\sl Numerical simulations with multiple-particle interactions,
from \protect{\cite{Gregoire04}} (dashed line in (a), (b) and (d))
and with binary collisions (full line). (a) average order parameter; (b)
Binder cumulant; (c) distribution of the order parameter for binary
collisions; (d) density profile along the coordinate $x_{\perp}$.
}
\label{fig-num}
\end{figure}

\textit{Binary versus multiple-particle interactions.}--
To confirm that binary collisions are sufficient to capture the
phenomena reported in numerical simulations \cite{Vicsek,Gregoire04},
we performed numerical simulations of a model with binary collisions
\footnote{For technical reasons, simulations were made
with a model slightly different (discrete time dynamics and non-Gaussian
noise) from the one we study analytically.},
and compared them with results obtained in a model
with multi-particle interactions \cite{Gregoire04}.
In both models, $N$ particles evolve on a periodic domain of
linear size $L$, with the same density $\rho_0=N/L^2=\frac{1}{8}$
in natural microscopic units ($L=256$ for the binary collisions model,
and $L=512$ for the other model).
The order parameter $\langle \varphi \rangle$, where
$\varphi=N^{-1}|\sum_{j=1}^N e^{i\theta_j}|$,
is shown on Fig.~\ref{fig-num}(a) as a function of the reduced
noise $\varepsilon = (\tilde{\sigma}_t-\tilde{\sigma})/\tilde{\sigma}_t$,
$\tilde{\sigma}^2$ being the variance of the noise,
and $\tilde{\sigma}_t$ the value of $\tilde{\sigma}$ at the transition.
Fig.~\ref{fig-num}(b) shows the Binder cumulant
$G=1-\langle \varphi^4 \rangle/3\langle \varphi^2 \rangle^2$.
The negative peak indicates a  discontinuous transition toward
spontaneous motion, which is confirmed in
Fig.~\ref{fig-num}(c) by plotting the probability distribution function
(PDF) of the order parameter (for binary collisions) for $\varepsilon$
below, above and very close to the transition
\footnote{The PDF is measured over a time interval of $500\, \tau$, where
$\tau$ is the correlation time.}. 
The distribution is clearly bimodal
at the transition, which is typical of discontinuous transitions.
Finally, Fig.~\ref{fig-num}(d) presents the density profile obtained
when spontaneous motion sets in, indicating the
presence of a stripe with higher density. This stripe is essentially
invariant along the $x_{\|}$ direction, and is moving along the
$x_{\perp}$ direction (on which the profile is measured, using a moving
frame). Note that the profile is asymmetric, with a higher slope on the front.
Thus, a model with only binary collisions is legitimate
and behaves qualitatively in a similar way as a model with more complicated
interactions.

\textit{Boltzmann equation.}--
The Boltzmann equation describing the evolution of the one-particle
phase-space distribution $f(\mathbf{r},\theta,t)$ reads
\be \label{eq-Boltz}
\frac{\partial f}{\partial t}(\mathbf{r},\theta,t) + \mathbf{e}(\theta)
\cdot \nabla f(\mathbf{r},\theta,t) = I_{\text{dif}}[f] + I_{\text{col}}[f]
\ee
where $I_{\text{dif}}[f]$ accounts for the self-diffusion phenomenon, and
$I_{\text{col}}[f]$ describes the effect of collisions; $\mathbf{e}(\theta)$
is the unit vector in the direction $\theta$.
$I_{\text{dif}}[f]$ is given by
\bea
I_{\text{dif}}[f] &=& -\lambda f(\mathbf{r},\theta,t)
+ \lambda \int_{-\pi}^{\pi} d\theta' \int_{-\infty}^{\infty} d\eta\,
p_0(\eta)\\
&& \quad \times
\sum_{m=-\infty}^{\infty} \delta(\theta'+\eta-\theta+2m\pi)
f(\mathbf{r},\theta',t)
\nonumber
\eea
The collision term $I_{\text{col}}[f]$ is evaluated as follows.
By definition, two particles collide if their relative distance becomes less
than $d_0$. In the referential of particle $1$, particle $2$ has a velocity
$\mathbf{v}_2'=\mathbf{e}(\theta_2)-\mathbf{e}(\theta_1)$. Thus,
particles that collide with particle $1$ between $t$ and $t+dt$ are those
that lie, at time $t$, in a rectangle of length $|\mathbf{v}_2'|$ and
of width $2d_0$. This leads to
\bea
I_{\text{col}}[f] &=&  -f(\mathbf{r},\theta,t) \int_{-\pi}^{\pi} d\theta'\,
|\mathbf{e}(\theta')-\mathbf{e}(\theta)| f(\mathbf{r},\theta',t)\\
\nonumber
&+& \int_{-\pi}^{\pi} d\theta_1 \int_{-\pi}^{\pi} d\theta_2
\int_{-\infty}^{\infty} d\eta\, p(\eta)\,
|\mathbf{e}(\theta_2)-\mathbf{e}(\theta_1)|\\
\nonumber
&\times& f(\mathbf{r},\theta_1,t) f(\mathbf{r},\theta_2,t)
\sum_{m=-\infty}^{\infty} \delta(\overline{\theta}+\eta-\theta+2m\pi)
\eea
with $\overline{\theta}=\text{Arg}(e^{i\theta_1}+e^{i\theta_2})$.
It can be checked easily that the uniform distribution
$f(\mathbf{r},\theta,t)=\rho/2\pi$, is a solution of
Eq.~(\ref{eq-Boltz}) for any density, and whatever the form of the noise
distributions $p_0(\eta)$ and $p(\eta)$.

\textit{Hydrodynamic equations.}--
Let us now define the hydrodynamic density and velocity fields
$\rho(\mathbf{r},t)$ and $\mathbf{u}(\mathbf{r},t)$
\bea
\rho(\mathbf{r},t) &=& \int_{-\pi}^{\pi} d\theta\, f(\mathbf{r},\theta,t)\\
\rho(\mathbf{r},t)\, \mathbf{u}(\mathbf{r},t) &=&  \int_{-\pi}^{\pi} d\theta\,
f(\mathbf{r},\theta,t)\,
\mathbf{e}(\theta)
\eea
Integrating the Boltzmann equation (\ref{eq-Boltz}) over $\theta$ yields
the continuity equation for $\rho(\mathbf{r},t)$
\be \label{continuity}
\frac{\partial \rho}{\partial t} + \nabla \cdot (\rho \mathbf{u}) = 0
\ee
The derivation of a hydrodynamic equation for the velocity field is less
straightforward, and involves an approximation scheme. Let us introduce
the Fourier series expansion of $f(\mathbf{r},\theta,t)$ with respect to
$\theta$
\be
\hat{f}_k(\mathbf{r},t) = \int_{-\pi}^{\pi} d\theta \,
f(\mathbf{r},\theta,t)\, e^{ik\theta}
\ee
Multiplying Eq.~(\ref{eq-Boltz}) by $e^{ik\theta}$ and integrating
over $\theta$ leads to an infinite set of coupled equations for
$\hat{f}_k(\mathbf{r},t)$.
We note that, identifying complex numbers with two-dimensional vectors
so that $e^{i\theta}$ corresponds to $\mathbf{e}(\theta)$, the Fourier
coefficient $\hat{f}_1(\mathbf{r},t)$ is nothing but the ``momentum'' field
$\mathbf{w}(\mathbf{r},t) = \rho(\mathbf{r},t)\, \mathbf{u}(\mathbf{r},t)$.
Thus the evolution equation for
$\hat{f}_1(\mathbf{r},t)$ should yield the hydrodynamic equation for
$\mathbf{u}(\mathbf{r},t)$. Yet, as $\hat{f}_k(\mathbf{r},t)$ is
coupled to all others $\hat{f}_l(\mathbf{r},t)$, a closure relation has
to be found.  In the following, we assume that the velocity distribution
$f(\mathbf{r},\theta,t)$ is only slightly non-isotropic,
or in other words that $\mathbf{u}(\mathbf{r},t)$
is small as compared to the individual velocity of particles,
and that the hydrodynamic fields vary on length scales that are much
larger than the microscopic length $d_0$.
As a result, the velocity equation is obtained from
the equation for $\hat{f}_1$ through an expansion to leading orders in
$\hat{f}_k(\mathbf{r},t)$ and in space and time derivatives.  Noting
that $\hat{f}_0(\mathbf{r},t)=\rho(\mathbf{r},t)=\mathcal{O}(1)$, we
set $\hat{f}_1(\mathbf{r},t)=\mathcal{O}(\epsilon)$,
$\epsilon \ll 1$. A consistent
scaling ansatz, confirmed by a numerical integration of
Eq.~(\ref{eq-Boltz}) in the steady state, is
$\hat{f}_k(\mathbf{r},t)=\mathcal{O}(\epsilon^{|k|})$.  Expanding to
order $\epsilon^3$, one only keeps the terms in $\hat{f}_1$ and
$\hat{f}_2$ in the evolution equation for $\hat{f}_1$. 
A similar expansion of the equation for $\hat{f}_2$ leads to
a closure relation for the equation on $\hat{f}_1$,
finally leading to the following hydrodynamic equation \cite{long}
\bea \label{eq-w}
\frac{\partial \mathbf{w}}{\partial t} + \gamma (\mathbf{w} \cdot \nabla)
\mathbf{w} &=& -\frac{1}{2}\nabla (\rho - \kappa \mathbf{w}^2) \\
\nonumber
&+& (\mu - \xi \mathbf{w}^2) \mathbf{w}
+ \nu \nabla^2 \mathbf{w} - \kappa (\nabla \cdot \mathbf{w}) \mathbf{w}
\eea
where the different coefficients are given by
\bea
\nu &=& \frac{1}{4} \left[ \lambda \left(1-e^{-2\sigma_0^2}\right)
+\frac{4}{\pi} \rho \left( \frac{14}{15}+\frac{2}{3}
e^{-2\sigma^2} \right) \right]^{-1}\\
\gamma &=& \frac{8\nu}{\pi} \left(\frac{16}{15}+2e^{-2\sigma^2}
-e^{-\sigma^2/2} \right)\\
\kappa &=& \frac{8\nu}{\pi} \left(\frac{4}{15}+2e^{-2\sigma^2}
+e^{-\sigma^2/2} \right)\\
\label{eq-mu}
\mu &=& \frac{4}{\pi}\rho \left(e^{-\sigma^2/2}-\frac{2}{3}\right)
-\lambda \left( 1-e^{-\sigma_0^2/2} \right)\\
\xi &=& \frac{64\nu}{\pi^2} \left( e^{-\sigma^2/2}-\frac{2}{5} \right)
\left( \frac{1}{3}+e^{-2\sigma^2} \right)
\eea
The first term in the r.h.s.~of Eq.~(\ref{eq-w})
may be thought of as a pressure gradient,
introducing an effective pressure $p=\frac{1}{2}(\rho-\kappa \mathbf{w}^2)$.
The second term describes the local relaxation of $\mathbf{w}$,
whereas the third term corresponds to the usual viscous term, and the last
one may be interpreted as a feedback from the compressibility of the flow.
The fact that $\gamma \ne 1$ (apart from special values of $\sigma$)
in the advection term expresses that the problem is not Galilean
invariant.
Note that $\nu$, $\gamma$ and $\kappa$ are always positive;
$\mu$ can change sign, and $\xi>0$ whenever $\mu>0$.
All the terms are compatible with the phenomenological
equation of motion of Toner \textit{et~al.} \cite{Toner}. However, our
approach provides explicit forms for the coefficients. In particular,
the coefficient in front of the term $\nabla (\nabla \cdot\mathbf{w})$ is
strictly zero in our case. Besides, there is no term of the form
${(\mathbf{w} \cdot \nabla)}^2 \mathbf{w}$ due to the order of truncation
of the Boltzmann equation.

\textit{Phase diagram.}--
We can now study the spontaneous onset of collective motion 
in the present model. As a first step, it is interesting to consider
possible instabilities of the spatially homogeneous system, that is
the appearance of a uniform nonzero field $\mathbf{w}$.
Equating all space derivatives to zero leads to the simple equation
\be \label{eq-homog}
\frac{\partial \mathbf{w}}{\partial t} = (\mu - \xi \mathbf{w}^2) \mathbf{w}
\ee
Clearly, $\mathbf{w}=0$ is solution for all values of the coefficients,
but it becomes unstable for $\mu>0$, when a nonzero solution
$\mathbf{w}_0= \sqrt{\mu/\xi} \, \mathbf{e}$ appears, where
$\mathbf{e}$ is an arbitrary unit vector. From Eq.~(\ref{eq-mu}), the
value $\mu=0$ corresponds to a threshold value $\rho_t$

\begin{figure}[t]
\centering\includegraphics[width=7.5cm,clip]{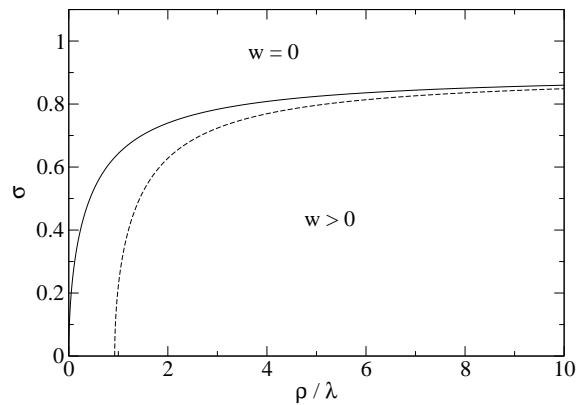}
\caption{\sl Phase diagram of the model in the plane $(\rho/\lambda,\sigma)$.
A transition line (full line: $\sigma_0=\sigma$; dashed line: $\sigma_0=1$)
indicates the linear instability threshold of the state
$w=|\mathbf{w}|=0$.
}
\label{phase-diag}
\end{figure}

\be
\rho_t = \frac{\pi \lambda (1-e^{-\sigma_0^2/2})}{4(e^{-\sigma^2/2}-
\frac{2}{3})}
\ee
The transition line defined by $\rho_t$ in the plane
$(\rho/\lambda,\sigma)$ is plotted on Fig.~\ref{phase-diag}, for
$\sigma_0=\sigma$ and for a fixed value $\sigma_0=1$. If $\sigma_0=\sigma$,
the instability occurs at any density, provided the noise is low
enough. On the contrary, at fixed $\sigma_0$, the instability disappears
below a finite density
$\rho_t^0=3\pi\lambda(1-e^{-\sigma_0^2/2})/4$.  Both transition lines 
saturate at a value $\sigma_t=(2\ln\frac{3}{2})^{1/2} \approx 0.90$.

Let us now test the stability against perturbations of the above spatially
homogeneous flow $\mathbf{w}(\mathbf{r},t)=\mathbf{w}_0$,
with finite density $\rho_0$, in an infinite space.
From Eq.~(\ref{eq-homog}), it is clear that
$\mathbf{w}_0$ is stable against spatially homogeneous perturbations.
Yet, this solution may be unstable against finite wavelength perturbations
\footnote{Note that for $\rho<\rho_t$, the solution $\mathbf{w}=0$ is stable
against finite wavelength perturbations \cite{long}.}.
To check this issue, we introduce a perturbation around the homogeneous
steady-state solution
\be
\rho(\mathbf{r},t) = \rho_0 + \delta \rho(\mathbf{r},t),
\quad
\mathbf{w}(\mathbf{r},t) = \mathbf{w}_0 + \delta \mathbf{w}(\mathbf{r},t)
\ee
and linearize Eq.~(\ref{eq-w}) in $\delta \rho(\mathbf{r},t)$ and
$\delta \mathbf{w}(\mathbf{r},t)$. Linear stability is then tested with
the ansatz
\be
\delta \rho(\mathbf{r},t) = \delta \rho_0\, e^{st+i\mathbf{q}\cdot\mathbf{r}},
\quad
\delta \mathbf{w}(\mathbf{r},t) = \delta \mathbf{w}_0 \,
e^{st+i\mathbf{q}\cdot\mathbf{r}},
\ee
where $\mathbf{q}$ is a given wavevector, by looking for the dispersion
relation $s(\mathbf{q})$.
By choosing $\delta \mathbf{w}_0$ and $\mathbf{q}$ along the same direction
as $\mathbf{w}_0$, one finds for the real part of $s$
\be
\Re(s) = \frac{\mu_0^2}{8\,\xi^3 w_0^4} \, |\mathbf{q}|^2 -
\frac{5\, \mu_0^4}{128\, \xi^7 w_0^{10}} \, |\mathbf{q}|^4
+ \mathcal{O}(|\mathbf{q}|^6)
\ee
for small $|\mathbf{q}|$,
with $\mu_0=4(e^{-\sigma^2/2}-\frac{2}{3})/\pi$ and $w_0=|\mathbf{w}_0|$,
indicating the onset of a long wavelength instability since $\Re(s)$ becomes
positive at small enough $|\mathbf{q}|$ \footnote{An analogous instability
has been reported in \cite{Ramaswamy}.}.
The spatially homogeneous states $\mathbf{w}=0$ and
$\mathbf{w}=\mathbf{w}_0$ are thus both unstable, so that more complicated
structures should eventually appear in the system.
The ``stripes'' of higher density moving over a low density
background, reported in \cite{Gregoire04}, may be examples of such patterns.
Physically, the instability may be interpreted as follows: if locally
$\rho>\rho_0$ ($\rho<\rho_0$), the local velocity $\mathbf{u}$ increases
(decreases), creating velocity gradients that generate a density wave.
Note that the perturbations that destabilize the long-range order
correspond to longitudinal waves, at odds with what happens in the
two-dimensional XY-model \cite{Kosterlitz} which might be thought of as an
equilibrium counterpart of the present model \cite{Toner}.

\textit{Discussion.}--
Our analytical approach has several advantages when compared with pure
numerical simulations of similar microscopic models. First, the hydrodynamic
equations may be used to get analytical solutions in reference cases
with simple geometries, and to analyze their stability against
perturbations.  Second, in more complicated situations, these equations
may be integrated numerically, allowing one to study much larger
systems than with direct simulations of the particles.

The hydrodynamic equations (\ref{continuity},\ref{eq-w})
have been derived from a
Boltzmann approach and their validity is in principle restricted to a low
density regime (note that the transition may occur at low density by choosing
$\lambda \ll 1$).
However, as verified for many systems, the validity of the Boltzmann
equation often goes well-beyond the a priori expected limit. One also
expects that in this low density regime, the hydrodynamic equations
should not depend strongly on the details of the interactions, as the
shape of $p(\eta)$. Another
limitation comes from the assumption that $\mathbf{w}$ is small.
This assumption is valid to describe the evolution of small perturbations
around the zero velocity state. When crossing the transition line,
the assumption is self-consistent, as the resulting homogeneous
velocity field grows continuously from zero.
Yet, the finite-wavelength instability may drive the
system into a regime where the approximation is no longer valid.
Checking this point requires to find the structures emerging from
Eqs.~(\ref{continuity},\ref{eq-w}), and to compare them with numerical
simulations. Work in this direction is under investigation \cite{long}.

\textit{Acknowledgments.}--
G.G. is grateful to H.~Chat\'e for introducing him to this subject.
This work has been partly supported by the Swiss National Science Foundation.

\end{document}